\documentclass[twocolumn,prb,showpacs,preprintnumbers]{revtex4}

\usepackage{graphicx}
\usepackage{dcolumn}
\usepackage{bm}
\usepackage{epsfig}

\begin{document}


\title{Thermodynamics and dynamics of the inherent states at the glass transition}

\author{U. Buchenau}
\affiliation{%
J\"ulich Center for Neutron Science, Forschungszentrum J\"ulich\\
Postfach 1913, D--52425 J\"ulich, Federal Republic of Germany\\
Corresponding author: Ulrich Buchenau,\\ email: buchenau-juelich@t-online.de 
}
\date{August 9, 2014}

\begin{abstract}
It has been recently shown that one can understand the Prigogine-Defay ratio at the glass transition in terms of freezing into one of the many inherent states of the undercooled liquid. In the present paper, the treatment is extended to the dynamics at the glass transition to show the connection to isomorphism and density scaling. In addition, the energy limits for stable inherent states are discussed.
\end{abstract}

\pacs{64.70.Pf, 77.22.Gm}

\maketitle

\section{Introduction}

When an undercooled liquid freezes into a glass at the glass temperature $T_g$, the sample stays in one of the many inherent states between which it could choose freely in the liquid \cite{palmer,stillinger,scio1,scio2,heuer}. Obviously, the possibility to jump from one possible stable structure to another gets lost as the system freezes into a glass, leaving only the vibrational degrees of freedom to contribute to dynamics and to thermodynamics.

Though this freezing process is in principle continuous, the explosive increase of the structural relaxation times with decreasing temperature at the glass temperature concentrates the freezing process into a relatively small temperature interval \cite{schmelzer}. The present paper is focused on the equilibrium liquid just above this temperature, in particular on the contributions of the possibility to jump from state to state to its thermodynamic and dynamic properties.

The concept of an inherent state stems from molecular dynamics simulations of undercooled liquids \cite{stillinger}, where one can remove the whole kinetic energy at a given moment in time and look for the nearest structural energy minimum of the atomic ensemble. The dynamic and thermodynamic consequences of the concept have been mainly studied in connection with numerical work \cite{scio1,scio2,heuer}, providing deep insights into the microscopic basis of the glass transition phenomenology. Instead of a single inherent state, it was found necessary to introduce the concept of a basin, an ensemble of inherent states with small barriers between them, in which the system stays a long time before jumping into the next basin.

Naturally, the concept is even better adapted to the glass transition in real systems, where the separation between vibrational and relaxational degrees of freedom is much more pronounced than in numerical simulations and the basin begins to resemble a single inherent state. Close to the glass transition, the system can then be considered to spend a long time vibrating in a single inherent state before it jumps into another one.

This picture has been recently applied \cite{prb2012} to the Prigogine-Defay ratio \cite{prigogine}, providing an explanation for values larger than one in terms of the properties of the inherent states. In the present paper, this treatment is extended to include predictions on the dynamics of the undercooled liquid. In addition, a consideration on the possible energy range of the stable inherent states in terms of melting and boiling points is given.

The paper gives a short summary of the theoretical basis and of the results of the preceding paper \cite{prb2012} in the next Section II following this introduction. The extension to dynamics follows in Section III. Section IV summarizes and concludes the paper.

\section{Properties of inherent states}

The inherent state is a structurally stable minimum of the potential energy for a sample of $N$ particles. $N$ should be large enough to get rid of finite size effects. A thermodynamic description of the undercooled liquid in terms of inherent states is reasonable in the temperature region where the structural relaxation is slow on the picosecond vibrational time scale, enabling one to distinguish between structural and vibrational entropy contributions.

An inherent state is characterized by its energy $Ne$ and its volume $Nv$ at the glass temperature $T_g$ and zero pressure, where $e$ is the average structural energy per particle and $v$ is the average particle volume. One has to specify temperature and pressure, because the volume of a given inherent state increases with temperature due to the vibrational anharmonicity and decreases with increasing pressure due to its compressibility.

With these definitions, it is clear that one of these inherent states has the lowest energy of the whole ensemble. That is the Kauzmann state. Its free energy must be higher than the one of the crystal; otherwise the liquid could never crystallize. However, from all the evidence at our disposal \cite{cavagna}, it seems clear that the free energy of this Kauzmann state is rather close to the crystalline one. The difference is probably negligible at the temperatures where one is able to study the undercooled liquid.

The Boltzmann factor for the inherent states contains not only the energy $Ne$, but its vibrational entropy $Ns_{vib}$ as well. One has to reckon with a vibrational entropy which depends on $e$. Denoting the average volume at the structural energy $e$ with $v_e$, it is reasonable to make a Gr\"uneisen Ansatz for the volume dependence of the vibrational entropy $s_{vib}$
\begin{equation}\label{svib}
	s_{vib}=s_{vib,K}+k_B\Gamma_e\ln{\frac{v_e}{v_K}},
\end{equation}
where $s_{vib,K}$ is the vibrational entropy of the Kauzmann state and $v_K$ is its atomic volume at $T_g$. As shown in the previous paper \cite{prb2012}, $\Gamma_e$ can be much larger than the usual \cite{kittel} Gr\"uneisen $\Gamma$, because it reflects the behavior of the boson peak rather than the one of the entire spectrum.

Here it is assumed that $s_{vib}$ depends {\it only} on the structural energy $e$ and not on the volume $v$. The assumption is necessary to keep the equations simple. It is supported by the finding that a vacancy in a crystal does practically not soften the vibrational spectrum \cite{scho}. One could argue that in two inherent states of equal energy, but different volume, the interatomic potential must be sampled at the same places, leading to a very similar vibrational spectrum. But it is an assumption which must not necessarily hold in every system.

The inherent state ensemble is described by its density in structural energy and volume at $T_g$ and zero pressure. Without loss of generality, one can split any distribution $g_{ev}(e,v)$ into a product
\begin{equation}\label{gv}
	g_{ev}(e,v)=g_e(e)g_v(e,v-v_e),
\end{equation}
with a normalized volume density at constant structural energy $g_v(e,v-v_e)$
\begin{equation}
	\int_0^\infty g_v(e,v-v_e)dv=1,
\end{equation}
an average volume $v_e$ at the structural energy $e$
\begin{equation}\label{v1}
	\int_0^\infty vg_v(e,v-v_e)dv=v_e,
\end{equation}
and a volume fluctuation contribution
\begin{equation}\label{v2}
	\int_0^\infty (v-v_e)^2g_v(e,v-v_e)dv=v_e^2.
\end{equation}
In the thermodynamic limit of large $N$, $g_e(e)$ and $g_v(e,v-v_e)$ are $N$-independent. In order to be able to work with the Boltzmann factor $\exp(-\beta Ne)$ alone, one defines the generalized distribution function
\begin{equation}\label{geg}
	g(e)=g_e(e)\left(\frac{v_e}{v_K}\right)^{\Gamma_e}.
\end{equation}

The average particle volume $v_e$ tends to increase with increasing structural energy $e$ due to the anharmonicity of the interatomic potential. This effect is responsible for the additional thermal expansion of the undercooled liquid. One assumes a linear relation
\begin{equation}\label{ve}
	v_e=v_K+a(e-e_K)
\end{equation}
where $e_K$ is the structural energy of the Kauzmann state and $v_K$ is its volume at the glass temperature $T_g$. The coefficient $a$, an inverse pressure, is a measure for the anharmonicity of the interatomic potential.

To get the partition function $Z$, one has to integrate the density $g_{ev}(e,v)$ of the inherent states per atom over the configurational energy $e$ and over the volume $v$. At zero pressure, the volume integrates out and one has
\begin{equation}\label{z}
	Z=\int_{-\infty}^\infty g(e)\exp(-\beta Ne)de,
\end{equation}
which contains the vibrational entropy contribution via eq. (\ref{geg}) for the generalized distribution function $g(e)$.

One can calculate the average structural energy $\overline{e}$ per atom and the average squared structural energy $\overline{e^2}$ per atom at zero pressure
\begin{equation}\label{ebar}
	\overline{e}=\frac{1}{Z}\int_{-\infty}^\infty eg(e)\exp(-\beta Ne)de
\end{equation}
and
\begin{equation}\label{e2bar}
	\overline{e^2}=\frac{1}{Z}\int_{-\infty}^\infty e^2g(e)\exp(-\beta Ne)de.
\end{equation}

The configurational part $\Delta c_p$ of the heat capacity at zero pressure per unit volume is given by
\begin{equation}\label{dcp}
	\Delta c_p=\frac{1}{v}\frac{\partial\overline{e}}{\partial T}=\frac{1}{vkT^2}(\overline{e^2}-\overline{e}^2).
\end{equation}

The average volume $\overline{v}$ is given by the double integral
\begin{equation}
\overline{v}=\frac{1}{Z}\int_\infty^\infty\int_0^\infty vg(e) g_v\exp(-\beta Ne)dedv.
\end{equation}
Because of eq. (\ref{v1}), one can again integrate the volume out and gets $\overline{v}=\overline{v_e}$, a single integral over $e$. Inserting eq. (\ref{ve}) for $v_e$, one gets
\begin{equation}\label{vav}
	\overline{v}=v_K+a(\overline{e}-e_K).
\end{equation}

The same procedure can be followed for the expectation value $\overline{v^2}$, this time using eq. (\ref{v2}) to evaluate the volume integral. Again replacing $v_e$ with eq. (\ref{ve}), one finds finally
\begin{equation}\label{central}
	\overline{v^2}-\overline{v}^2=a^2(\overline{e^2}-\overline{e}^2)+\overline{v_e^2},
\end{equation}
where $\overline{v_e^2}$ is the thermal average over the values $v_e^2$ at the different structural energies.

This central result \cite{prb2012} shows that one has two kinds of density fluctuations in the undercooled liquid. Those in the first term of the right side of eq. (\ref{central}) stem from a change of the structural energy, those in the second term occur at constant structural energy. Of course, in a given transition from one inherent state to another one will usually find a mixture of both. But there are indeed substances where the second term is practically zero, which implies a strong correlation between energy and density fluctuations, the property of isomorphism which is actively debated in the community \cite{nick,ulfth,gundermann,iso,isonew}.

Returning to the additional thermal expansion at $T_g$, one finds from eq. (\ref{vav}) 
\begin{equation}\label{a}
	\Delta\alpha=a\frac{1}{v}\frac{\partial\overline{e}}{\partial T}=a\Delta c_p.
\end{equation}

From the volume fluctuations, one calculates the additional compressibility $\Delta\kappa$ as in eq. (\ref{dcp}) 
\begin{equation}
\Delta\kappa=\frac{\overline{v^2}-\overline{v}^2}{Vk_BT}.
\end{equation} 

The two terms in eq. (\ref{central}) give rise to two terms in $\Delta\kappa$
\begin{equation}
	\Delta\kappa=\Delta\kappa_{PD}+\Delta\kappa_0
\end{equation}

The first is the compressibility contribution from the possibility to change the average structural energy
\begin{equation}\label{dk}
	\Delta\kappa_{PD}=a\frac{1}{v}\frac{\partial\overline{e}}{\partial p}=\frac{a^2}{VkT}(\overline{e^2}-\overline{e}^2).
\end{equation}
This part of the compressibility has the index $PD$, because it satisfies the Prigogine-Defay relation for a second-order phase transition \cite{jackle}
\begin{equation}\label{prigo}
	\frac{\Delta c_p\Delta\kappa_{PD}}{(\Delta\alpha)^2T}=\frac{\overline{\Delta E^2}\ \ \overline{\Delta V^2}}{(\overline{\Delta E\Delta V})^2}=1.
\end{equation}
Here $\Delta E$ and $\Delta V$ are the additional energy (enthalpy) and volume fluctuations from the structural energy changes, respectively. For completely correlated enthalpy and volume fluctuations (an implicit property of the first term), the Prigogine-Defay ratio is one.

But there is also the second term
\begin{equation}
	\Delta\kappa_0=\frac{\overline{v_e^2}}{vk_BT},
\end{equation}
which stems from the density fluctuations at constant structural energy. At zero pressure, these additional density fluctuations occur at constant energy and do neither contribute to the heat capacity nor to the thermal expansion. As a consequence, one finds the Prigogine-Defay ratio
\begin{equation}\label{PD1}
	\Pi=\frac{\Delta\kappa_0+\Delta\kappa_{PD}}{\Delta\kappa_{PD}}=\frac{\Delta\kappa}{\Delta\kappa_{PD}},
\end{equation}
where $\Delta\kappa$ is the measured value. $\Pi$ is larger than one if $\Delta\kappa_0$ is larger than zero. This provides a new explanation of the much-debated \cite{prigogine,davies,gupta,jackle,schmelzer} deviation of the Prigogine-Defay ratio from one at the glass transition.

One must bear in mind, however, that one has an important class of glass formers, the strongly correlated Roskilde liquids \cite{isonew}, in which $g_v(e,v)$ is close to a $\delta$-function, its equilibration plays no role and the Prigogine-Defay ratio is close to 1. They occur more often on the computer, but they are also found in reality \cite{gundermann}. They show strikingly simple properties, both in dynamics \cite{ellegaard} and thermodynamics \cite{iso,isonew}. 

After this summary of the previous paper \cite{prb2012}, let us first discuss a minor point, the natural limitations of $g(e)$ in the structural energy. As already emphasized above, the low energy limit is the Kauzmann state, which one cannot reach, but which is probably in energy and vibrational entropy close to the crystal \cite{cavagna}. The high energy limit of $g(e)$ is given by $k_BT_b$, where $T_b$ is the boiling point. Above this point, one does no longer expect stable structures.

Note that this consideration gives a very different energy range to different substances. In the noble gases, the boiling point lies very close to the melting point, so the range for stable structures is small. In the alkali metals, the boiling point lies a factor of three higher than the melting point, allowing for a rather large range of stable structures in the structural energy. The Lennard-Jones potential has a phase diagram which is very similar to the one of the noble gases, so it also has a relatively small range of stable structures.

The situation is simple if one has a gaussian density of states in the structural energy, according to numerical simulation \cite{heuer} the generic case. For $g(e)=g_0\exp\left(\frac{-(e-e_0)^2}{2w^2}\right)$, one calculates \cite{prb2012} an excess heat capacity
\begin{equation}\label{cpt}
	\Delta c_p=\frac{w^2}{Vk_BT^2}.
\end{equation}
Measured excess heat capacities per atom at $T_g$ lie between 0.3 and 2 $k_B$, so the width $w$ of the distribution is of the order of $k_BT_g$. For $w=k_BT_g$, the heat capacity per atom is exactly $k_B$. The full width at half maximum of the gaussian is $w\sqrt{8\ln2}$, so the whole gaussian requires an energy range of about $5w$. 

At atmospheric pressure, the lower energy limit of the stable inherent states is $k_BT_K$ ($T_K$ Kauzmann temperature, which usually lies around 0.6$T_g$ \cite{angell}) and the upper energy limit is estimated to lie around $k_BT_b$, so $5w\approx k_B(T_b-T_K)$. If the boiling point lies close to the melting point $T_m\approx T_g/0.6$, like, for instance, in the Lennard-Jones system, the whole gaussian cannot be much broader than about $1.5k_BT_g$, bringing $w$ down to $0.3k_BT_g$ and the excess heat capacity down to about $0.1k_B$ per atom. In fact, the excess heat capacity determined numerically at constant volume for the Lennard-Jones system \cite{ulfth} is only 0.15 $k_B$, consistent with a rather narrow distribution in the structural energy.

After this general consideration on the energy limits, let us proceed to the treatment of the dynamics. 

\section{Dynamics of inherent states}

Consider two inherent states of an isomorph glass former with structural energies $E_1$ and $E_2$. At $T_g$ and zero pressure, they have the weighted energy difference $(E_2-E_1)/k_BT_g$.

Applying a small pressure $p$, the weighted energy difference increases by the factor $(1+pa)$. To restore it to the original value, one has to raise the temperature by $\Delta T=paT_g$. Then one returns to the same weighted energy difference between the two states, keeping their thermal occupation ratio constant. If this holds for any combination of states, one returns to the same state population as the initial state at pressure zero - one moves on an isomorph \cite{iso,isonew}.

The question is: What happens to the structural relaxation time $\tau_\alpha$ on this isomorph $T(p)=T_g+pa$? Answer: It remains constant if the energy barrier between any two states scales also with the factor $1+pa$, because then the relaxation time for the thermally activated jumps between them stays the same. If this holds for any two states, the structural relaxation time $\tau_\alpha$ at $p$ and $T_g+pa$ is the same as the zero pressure one at $T_g$. This implies that the isomorph is also an isochrone, a curve with a constant structural relaxation time.

In a truly isomorph glass former, this equality holds by definition. But note that now one strains the definition much more, because in order to satisfy the dynamic requirement one has to include the saddle points of the energy landscape into the same definition, about 30 $k_BT_g$ higher in energy than the minima. Thus one cannot help wondering whether real glass formers at $T_g$ will indeed exhibit isomorphism.

But this can be checked. If minima and saddle points lie on the same $e(v)$ curve, moving on $T(p)=T_g+pa$ does not change the relaxation time. Then the pressure dependence of $T_g$ is given by
\begin{equation}\label{ehrenfest}
	\frac{\partial T_g}{\partial p}=aT_g=\frac{\Delta\alpha T_g}{\Delta c_p}.
\end{equation}
Note that the second part of this equation is one of the two Ehrenfest equations (the one connected with the entropy) for the pressure dependence of a second order phase transition \cite{prigogine,schmelzer}. Here it is for the first time derived in an appropriate way for a glass former. As it turns out, it is very often reasonably well fulfilled at the glass transition \cite{angell,reilly}, not only for glass formers with a Prigogine-Defay ratio close to one, but also for glass formers with a high Prigogine-Defay ratio like glycerol and B$_2$O$_3$. But it is not always fulfilled: polyvinylchloride, for example, shows a $\frac{\partial T_g}{\partial p}$ which is nearly a factor of two smaller than the Ehrenfest equation eq. (\ref{ehrenfest}). A possible explanation is: In this case the saddle points are a factor of two less pressure-sensitive than the structural minima. But there is a second possible explanation to be discussed below, which is more probable.

Next question: How is it possible that nonisomorphic substances obey the isomorph law eq. (\ref{ehrenfest})? The answer is simple: According to the derivations of the preceding section, non-isomorph or non-Roskilde substances are characterized by non-zero fluctuations $v_e^2$ of the volume at constant structural energy $e$. Now let $v_e^2$ be independent of $e$. Then the above scaling arguments hold again for the average volume difference between two states at $E_1$ and $E_2$. If again the average saddle point lies on the same $e(v)$-curve as the average of the minima, eq. (\ref{ehrenfest}) should hold.

But the assumption of an energy-independent $v_e^2$ is not the natural one. One would rather guess that $v_e^2$ increases with $e$, because $g(e)$ itself increases with $e$ in the relevant region \cite{scio1,scio2,heuer}. Qualitatively, such an increase must lead to a smaller $\partial T_g/\partial p$ than the one of the Ehrenfest eq. (\ref{ehrenfest}), because the tendency to shift the states to smaller $e$ by an external pressure is diminished. This second explanation leads to a smaller $\partial T_g/\partial p$ than the entropy Ehrenfest equation (\ref{ehrenfest}) like in polyvinylchloride and in several other examples \cite{angell,reilly,roland}.  

\begin{table}[htbp]
	\centering
		\begin{tabular}{|c|c|c|c|c|c|c|}
\hline
substance          & $T_g$ & $\Delta c_p$ & $\Delta\alpha$ & $\kappa$ &  $\gamma_{exp}$ & $\gamma_{th}$ \\
\hline   
                   & K   &  10$^6$J/m$^3$ & 10$^{-4}$ K$^{-1}$ & 1/GPa &       &            \\
\hline   
glycerol           & 183 &   1.16   &      3.9           &   0.18&   1.8   & 2.2  \\
OTP                & 236 &   0.60   &      5.7           &   0.34&   4.0   & 4.4  \\
PVAc               & 304 &   0.59   &      4.3           &   0.50&   2.6 1.4 & 2.4\\
PMMA               & 378 &   0.35   &      3.5           &   0.58&   1.25  & 2.2  \\
\hline		
		\end{tabular}
	\caption{Comparison of measured \cite{roland} density scaling exponents $\gamma_{exp}$ with values $\gamma_{th}$ calculated via eq. (\ref{g}) from the thermodynamic data at the glass transition \cite{jcp2012}}.
	\label{tab:Comp}
\end{table}

If the Ehrenfest relation of eq. (\ref{ehrenfest}) holds, one can proceed to calculate the density scaling exponent $\gamma$ of the density scaling relation \cite{dreyfus,christiane,roland} for the structural relaxation time $\tau_\alpha$
\begin{equation}\label{gamma}
	\tau_\alpha=f(TV^\gamma).
\end{equation}

If the system stays at constant structural energy, the Prigogine-Defay contribution to the compressibility vanishes. Then the compressibility on the isochrone is
\begin{equation}\label{compi}
	-\frac{1}{V}\frac{\partial V}{\partial p}\mid_\tau=\kappa-\Delta\kappa_{PD}.
\end{equation}
Since
\begin{equation}
\frac{\partial V}{\partial T}\mid_\tau=\frac{\partial V}{\partial p}\mid_\tau\frac{\partial p}{\partial T}\mid_\tau=\frac{1}{aT_g}\frac{\partial V}{\partial p}\mid_\tau
\end{equation}
one gets the equation for the density scaling exponent $\gamma$
\begin{equation}\label{g}
	\frac{1}{\gamma}=-\frac{T_g}{V_g}\frac{\partial V}{\partial T}\mid_\tau=\frac{\Delta c_p\kappa}{\Delta\alpha}-\Delta\alpha T_g
\end{equation}
where eq. (\ref{a}) has been used to express $a$ in terms of measurable quantities and eq. (\ref{prigo}) has been used to replace $\Delta\kappa_{PD}$.

In order to check this equation, one needs substances where both $\gamma$ and the Prigogine-Defay ratio have been determined. At present, there are only the four examples in Table I for which both measurements have been done. For glycerol, orthoterphenyl and polyvinylacetate, one finds agreement within experimental error (note the two different $\gamma$-values of polyvinylacatate obtained in two independent measurements). For polymethylmethacrylate, the agreement is poor; however, this is again a case where the measured \cite{roland} $\frac{\partial T_g}{\partial p}=240$ K/GPa is a factor 1.57 lower than the value of 377 K/GPa calculated from the data in Table I with the entropy Ehrenfest equation (\ref{ehrenfest}), so one cannot expect good agreement.

Very recently, Casalini and Roland \cite{roland2} have independently derived a very similar equation for $\gamma$, which in the present formulation reads
\begin{equation}\label{gr}
\frac{1}{\gamma}=\frac{\Delta c_p\kappa}{\Delta\alpha}-\alpha T_g,	
\end{equation}
and only differs in the second term, where the $\Delta\alpha$ of eq. (\ref{g}) is replaced by the full $\alpha$, so one subtracts a term which is about a factor of 3/2 larger and gets a larger $\gamma$. Within the large experimental error, it is difficult to tell from experiment which of the two equations is better.

One can trace the difference between the two equations back to their different derivations. The equation of Casalini and Roland \cite{roland2} starts from the equation between the activation energy $E_V$ at constant volume and $E_p$ at constant pressure \cite{roland2004}
\begin{equation}
	\frac{E_V}{E_p}=\frac{1}{1+\alpha\gamma T_g},
\end{equation}
which does not distinguish between the vibrational anharmonicity expansion and the expansion due to the occupation of structures with higher energy. For this reason, the full thermal expansion $\alpha$ enters into eq. (\ref{gr}). In the present work, where the two mechanisms are distinguished, one finds $\Delta\alpha$ (the expansion due to structural changes only) in the equation.

\section{Summary}

To summarize, the concept of inherent states does not only give insight into the physics of the Prigogine-Defay ratio, but also allows to make quantitative predictions for the dynamics. These are not first-principle predictions, because they require the assumption that the vibrational entropy of an inherent state is exclusively given by its energy and is independent of its volume. But they allow for a quantitative analysis of the low temperature undercooled liquid at the real glass transition, which is not accessible to numerical methods. Also, they allow for a badly needed physical insight into the mechanisms of the glass transition. On the basis of the scheme, one predicts a small difference between the heat capacities of glass and undercooled liquid if the melting point and the boiling point lie close together, because then there is no room for a broad distribution of stable inherent states.

The formalism allows for the first time an appropriate glass-former derivation of the entropy Ehrenfest relation which has been found to be often valid. Its validity is not due to the resemblance of the glass transition to a second order phase transition, but is rather a consequence of a linear relation between structural energy and volume between the inherent states, which continues to remain valid for the saddle points of the energy landscape needed for the structural relaxation. This condition can hold or not hold, independent of whether the system shows isomorphism or not; if the system is non-isomorph (not a Roskilde liquid), the validity of the Ehrenfest relation requires not only the scaling behavior of the minima and the saddle points, but in addition an energy independence of the volume fluctuations at constant energy. If the volume fluctuations at constant structural energy increase with the structural energy, one gets a smaller pressure dependence than the one predicted by the entropy Ehrenfest relation.  

If the entropy Ehrenfest relation holds, one can derive an equation for the calculation of the density scaling exponent $\gamma$ from measurable physical quantities. The equation differs slightly from another recently derived one, because its derivation distinguishes between two mechanisms of thermal expansion, the influence of the vibrational anharmonicity and the population change of the inherent states.

\end{document}